\documentclass[useAMS,usenatbib]{mnras}
\usepackage{times,graphicx,amsmath,amsfonts,amssymb,aas_macros,epstopdf,ulem}
\usepackage{epsfig}
\usepackage[usenames,dvipsnames]{xcolor}
\usepackage{url}
\usepackage{subfigure}

\def\etal{{\frenchspacing\it et al.}}
\def\ie{{\frenchspacing\it i.e.}}
\def\eg{{\frenchspacing\it e.g.}}

\def\be{\begin{equation}}
\def\ee{\end{equation}}
\def\ba{\begin{eqnarray}}
\def\ea{\end{eqnarray}}

\def\msol{{\rm M}_{\odot}}
\def\hmpc{h^{-1}\,{\rm Mpc}}

\def\hkpc{h^{-1}\,{\rm kpc}}
\def\mpch{h\,{\rm Mpc}^{-1}}

\def\frac#1#2{{\textstyle{#1\over #2}}}

\def\simlt{\stackrel{<}{{}_\sim}}
\def\simgt{\stackrel{>}{{}_\sim}}

\newcommand{\eagle}{\textsc{eagle}}
\newcommand{\dmonly}{\textsc{dmo}}

\newcommand{\dtfe}{\textsc{dtfe}}
\newcommand{\lcdm}{$\Lambda$CDM~}

\title[DM simulations and galaxy formation]{The effect of baryons on redshift space distortions and cosmic density and velocity fields
in the EAGLE simulation}
\author[Wojciech A.~Hellwing \etal]{Wojciech A. Hellwing$^{1,2,3}$\thanks{E-mail: pchela@icm.edu.pl}, 
Matthieu Schaller$^{2}$, Carlos S. Frenk$^{2}$, Tom Theuns$^{2}$,\newauthor 
Joop Schaye$^{4}$, Richard G.~Bower$^{2}$, Robert A.~Crain$^{5}$\\
$^{1}$Institute of Cosmology and Gravitation, University of Portsmouth, Portsmouth PO1 3FX, UK\\
$^{2}$Institute for Computational Cosmology, Department of Physics, Durham University, South Road, Durham DH1 3LE, UK\\
$^{3}$Janusz Gil Institute of Astronomy, University of Zielona G\'ora, ul. Lubuska 2, Zielona G\'ora, Poland\\
$^{4}$Leiden Observatory, Leiden University, PO Box 9513, NL-2300 RA Leiden, the Netherlands\\
$^{5}$Astrophysics Research Institute, Liverpool John Moores University, 146 Brownlow Hill, Liverpool L3 5RF, UK\\
}
\begin{document}

\date{Accepted XXXX . Received XXXX; in original form XXXX}

\pagerange{\pageref{firstpage}--\pageref{lastpage}} \pubyear{2016}

\maketitle

\label{firstpage}

\begin{abstract}
We use the EAGLE galaxy formation simulation
to study the effects of baryons on the power spectrum of the total matter
and dark matter distributions and on the velocity fields of dark matter and galaxies.
On scales $k\simgt 4\mpch$ the effect of baryons on the amplitude 
of the total matter power spectrum is greater than
$1\%$. The back-reaction of baryons 
affects the density field of the dark matter at the level of $\sim3\%$ on scales
of $1\leq k/(\mpch)\leq 5$. The dark matter velocity divergence
power spectrum at $k\simlt0.5\mpch$ is changed by less than $1\%$. 
The 2D redshift-space power 
spectrum is affected at the level of $\sim6\%$ at $|\vec{k}|\simgt 1\mpch$
(for $\mu>0.5$), but for $|\vec{k}|\leq 0.4\mpch$ it differs by less than $1\%$.
We report vanishingly small baryonic velocity bias for haloes:
the peculiar velocities of haloes with $M_{200}>3\times10^{11}\msol$ 
(hosting galaxies with $M_{*}>10^9\msol$) are affected at the level of at most 
$1~$km/s, which is negligible for $1\%$-precision cosmology.
We caution that since EAGLE overestimates cluster gas fractions
it may also underestimate the impact of baryons, 
particularly for the total matter power spectrum.
Nevertheless, our findings suggest that for theoretical modelling of 
redshift space distortions and galaxy velocity-based statistics,
baryons and their back-reaction can be safely ignored at the current
level of observational accuracy. 
However, we confirm that the modelling
of the total matter power spectrum in weak lensing studies 
needs to include realistic galaxy formation physics
in order to achieve the accuracy required in the precision cosmology era.  
\end{abstract}


\section{Introduction}
\label{sec:intro}
The standard hierarchical structure formation theory assumes that 
the distribution of mass in the Universe has evolved out of primordial 
post-inflationary Gaussian density and velocity perturbations via 
gravitational instability. The resulting large-scale structures 
can be described in a statistical way. Two-point statistics 
(power spectrum and correlation function) are the most 
widely studied measures 
\citep[see \eg][]{1980Peebles,cl_stat_dyn,Percival2001,Gaztanaga2002,Cole2005,Eisenstein2005}. 
With the advent of precision cosmology, defined here as a level of one percent precision in cosmic observables,
it is a matter of utmost relevance to obtain accurate 
theoretical estimates of the two-point statistics. Theoretical
modelling is needed to assess and model the systematic effects
present in cosmic observables. This modelling needs
to be precise enough to reduce the impact of
the systematic effects below that of the expected statistical errors.
So far the common approach has been to use large computer N-body simulations
of a collisionsless dark matter (DM) fluid 
\citep[see \eg][for an extensive review]{FrenkWhite2012},
to model the cosmic density and velocity fields. DM-only simulations are relatively simple
and cheap in terms of computer resources. However, they treat the baryonic component
in a simplified manner, modelling it as dark and pressureless. 
In the light of the accuracy required
by precision cosmology this approach might well turn out to be
inadequate for accurate modelling of all relevant systematic effects.

In linear theory baryons follow the gravitational evolution of
dark matter, which dominates the gravitational potential 
on large scales (\ie~tens of Megaparsecs). 
However, on smaller scales the highly non-linear
nature of the physical processes that govern galaxy formation can lead to
significant displacement of the baryonic components relative to the
underlying DM 
\citep[\eg][]{Jing2006,Rudd2008,Guillet2010,vanDaalen2011,Velliscig2014,vanDaalen2014,Mohammed2014}. 
On those smaller scales, we
can distinguish two different regimes. The first one concerns
scales to hundreds of kiloparsecs, where owing to radiative cooling,
gravitationally preheated gas can efficiently dissipate internal
energy and condense into halo centres reaching densities much
higher than those of the accompanying DM. This effect boosts the
variance of the baryon density field w.r.t that of the DM by 10-20\% on scales
$<500 \hkpc$ \citep[\eg][]{vanDaalen2011}, (hereafter VD11).  
The second one is connected to the very energetic processes of {\it Supernovae}
(SN) explosions and other stellar feedback events, 
as well as feedback from {\it Active Galactic Nuclei} (AGN).
These feedback processes can eject significant amounts of gas from the galaxies
and haloes in which they reside. Especially efficient AGN energy
feedback leads to expulsion of gas from 
the high-redshift progenitors of today's group and cluster
sized haloes beyond their $z=0$ virial radii. 
Simulations require such energetic feedback to 
match simultaneously optical and X-ray observations of galaxy groups and clusters 
\citep[\eg][]{McCarthy2010,Fabjan2010,McCarthy2011}.
Hence SN and AGN feedback yield smoother baryon density
contrasts on scales up to a few Megaparsecs \citep[\eg VD11;][]{Puchwein2013,Vogelsberger2014}. 
  
We expect that on small and intermediate scales (\ie $\simlt 20\hmpc$),
the distribution of baryonic matter could differ significantly 
from that of the collisonless component and that this will produce a back-reaction
onto the dark matter (\eg VD11). This back-reaction, in turn, can produce non-negligible
effects in the DM distribution on galactic and intergalactic scales.
The baryonic back-reaction may also affect the velocity fields of DM, haloes and galaxies. 
While these baryonic effects
on the total and dark matter density fields have been studied in previous works,
the impact on the cosmic velocity fields of DM and 
galaxies remains to be investigated. Accurate modelling
of this phenomenon is important 
since extraction of cosmological information
from galaxy redshift surveys requires precise modelling of the galaxy 
and DM peculiar velocity fields. 

Our aim in this study is to assess the scale and size of the baryonic back-reaction on
both the cosmic density and velocity fields of DM and galaxies.
We will do this by analysing the state-of-the-art galaxy formation simulation
\eagle~(\citet{Eagle} hereafter S15, \citet{Crain2015}).

\section{The EAGLE simulation suite}
\label{sec:simulation}

In this letter, we use the main simulation (Ref-L100N1504, hereafter
\eagle) of the \eagle~({\it Evolution and Assembly of GaLaxies and their Environments}, S15)
~suite and its DM-only version (hereafter
\dmonly) that was run from the same initial conditions. This
was achieved by increasing the DM particle mass by a factor of
$(1-f_b)^{-1}$ (here $f_b\equiv\Omega_b/\Omega_{m}$ is the universal
baryon fraction). \eagle~uses a
state-of-the-art set of subgrid models and treatment of smoothed
particle hydrodynamics. 
The simulations assumes a flat \lcdm cosmology with parameters
from Planck2013 \citep{Planck2013}.  The initial conditions are
generated at $z=127$ using second-order Lagrangian perturbation theory
in a $100^3~\rm{Mpc}^3$ volume with a DM particle mass of
$9.7\times10^6~\msol$ and initial gas particle mass of
$1.8\times10^6~\msol$ \citep{Panphasia}. The particles are then evolved in time using a
modified version of the \textsc{Gadget} Tree-SPH code
\citep{Springel2005} that includes the pressure-entropy formulation of
the SPH equations by \cite{Hopkins2013} and other improvements whose
effects on the resulting galaxy population are discussed by
\cite{Schaller2015b}. 
The maximum physical Plummer-equivalent gravitational softening
is $\epsilon=700~\rm{pc}$.

The subgrid model in this simulation includes element-by-element
radiative cooling \citep{Wiersma2009a}, a star formation recipe
designed to reproduce the observed Kennicutt-Schmidt relation
\citep{Schaye2008}, chemical enrichment via stellar
mass loss \citep{Wiersma2009b}, stellar feedback
\citep{DallaVecchia2012}, gas accretion onto supermassive black holes
and the corresponding AGN feedback \citep{Booth2009,RosasGuevara2013}.
The simulation has been shown to reproduce broadly a variety of other
observables \citep[for details see S15;][]{Lagos2015,Bahe2016,Furlong2015,Rahmati2015,Schaller2015a,Trayford2015}.
With all these successes it is worth mentioning here also a significant
shortcoming of the simulation. The \eagle{} X-ray properties of groups and clusters presented
in S15 compares rather poorly with observations, with \eagle{}
predicting too high gas fractions in those objects. 
While S15 have shown that \eagle{} model 
AGNdT9 (which uses more efficient AGN feedback)
does much better, its box size of 50 Mpc is too small for our purposes.
This discrepancy is important 
in assessing the prominence of the baryonic effects at intergalactic scales,
as the gas fraction of massive objects is a sensitive tell-tale sign of the strength 
of baryonic effects on the corresponding scales \citep[see \eg][]{Semboloni2011,Sembolini2013}.
This should be borne in mind when we analyse the magnitude and scales of
the bayonic effects onto the matter spectrum in the \eagle{} simulation.

\section{Baryonic effects}
\label{sec:spectra}

We consider basic two-point statistics of the cosmic density and
velocity fields in the form of power spectra.  Specifically, we examine
the real-space total and DM power spectra of density fluctuations,
$P(k)\equiv\langle \delta_k\delta^*_k\rangle$,
the power spectrum of the scaled velocity divergence (expansion scalar),
$P_{\theta\theta}(k)\equiv\langle \theta_k\theta^*_k\rangle$,
defined here as $\theta_k\equiv {\nabla\cdot \vec{v}(\vec{k})/(aHf)}$.
The corresponding density-velocity cross-power spectrum is
$P_{\delta\theta}(k)\equiv\langle \delta_k\theta^*_k\rangle$,
and the full two-dimensional redshift space density power spectrum is
$P^s(k_\perp,k_\parallel)=\sum_{l=0}^\infty
P^s_l(|\vec{k}|)\mathcal{P}_l(\mu),$
with monopole moment $P_0^s(k)_{l=0}$, and quadrupole moment $P_2^s(k)_{l=2}$.
Here $\vec{k}$ is the comoving 3D Fourier mode wavevector, 
$\mu=\cos(|\vec{k}|/k_\parallel)$, $\vec{v}$ is the peculiar
velocity, $a$ is the cosmic scale factor, $H$ is the Hubble parameter,
$f$ is the growth rate of density fluctuations (defined as the logarithmic derivative
of the density perturbation growing mode with respect to the scale factor), and finally
$\mathcal{P}_l$ are Legendre polynomials.  For all calculations in
redshift space we use the distant observer approximation in which the
$z$-axis of the simulation cube is parallel to the observer's line of
sight ($\parallel$-direction) and the $x,y$-axes form a plane
perpendicular to the observer's direction ($\perp$-direction).
To compute the power spectra, we estimate the density and velocity fields
using the {\it Delaunay Tesselation Field Estimator} method of \citet{sv2000},
implemented in the publicly available ~code by \citet{cv2011}.
The \dtfe~method gives a volume-weighted velocity field and has a self-adaptive
smoothing kernel that follows the local density of tracers. 

For 1D spectra we
sample the fields onto a $1024^3$ cubic grid and for $3D$ spectra we
use a $512^3$ sampling grid. The size of the sampling grids implies
Nyquist limits for the spectra of $k_{Nyq}^{1024}=48.2\mpch$ and
$k_{Nyq}^{512}=24.1\mpch$ respectively. The analysis of lower-resolution
runs of \eagle~ indicates that the power spectra are converged to
$1\%$ at $k_{Nyq}/8$. However, since we are focused here on relative
differences between \dmonly{} and \eagle, we will consider the power spectra up
to their respective Nyquist sampling limits.

\begin{figure} 
  \includegraphics[angle=0,width=78mm]{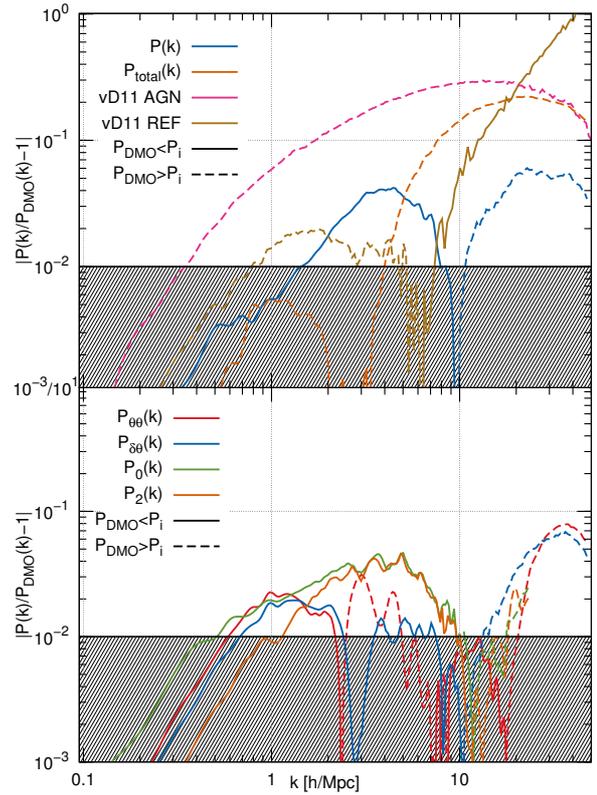}
   \caption{The relative difference of various power spectra in \eagle{}
    w.r.t the \dmonly{} case. 
    {\it Both panels}:
    The regime corresponding to baryonic corrections smaller than $1\%$
    is indicated as the hashed area. 
    Whenever the \eagle{} base power spectrum
    has a larger amplitude than \dmonly{} we use solid lines;
    in the opposite case we used dashed lines.
    {\it Top panel}:
    The blue line depicts the dark matter power spectrum, 
    the orange line illustrates the total matter $P(k)$.
    We also plot results for the total power spectra in two
    OWLS models (VD11): AGN (magenta line) and REF (tan line).
    {\it Bottom panel}:
    The green (orange) line shows the dark matter monopole (quadrupole) redshift
    space power spectrum, is the quadrupole, the red line the velocity divergence power
    spectrum, and the blue line the density-velocity cross-power spectrum.
     }
 \label{fig:pk_compare}
\end{figure}

In Fig.~\ref{fig:pk_compare} we plot all relevant \eagle{} one-dimensional power
spectra as absolute values of their relative
differences with respect to the corresponding \dmonly~ power spectra. For all
cases the dashed lines mark the results when the \eagle{}
amplitude is lower than the \dmonly{} case, whilst the solid lines correspond to the opposite.
We first focus on the total matter power spectrum (orange line). 
Theoretical predictions of this statistic up to $k\sim 5\mpch$ are
needed for precision cosmology with upcoming 
surveys like Euclid \citep{Laureijs2011} and LSST~\citep{Ivezic2008}.  
The simulation suggests that at $k=5\mpch$ baryons already produce a
$5\%$ difference in the amplitude. This effect is much more
pronounced when we consider even smaller scales: at
$k\sim10-20\mpch$ the difference between \dmonly~and \eagle~can
be as large as $10-20\%$. 
The results are compared with two of the OWLS models
\citep{Schaye2010} analysed by VD11.
Our results for $k\geq 5\hmpc$
fall in between VD11 REF model (which had
no AGN feedback; tan line) and their AGN model (with strong AGN
feedback; magenta). However, at larger scales, we observe that
the effect seen in \eagle{} is weaker than their REF model.
This regime is affected by \eagle{} limited volume
\footnote{\ie{} relative lack of extreme objects like rich clusters},
and thus susceptible to cosmic variance. 

We evaluate the amplitude and scales on which
the back-reaction of baryons affects the DM by studying the blue
line in Fig.~\ref{fig:pk_compare}, which shows that on scales
$k>5\mpch$ the back-reaction effects are much smaller (up to $6\%$)
than the baryonic effects we have seen in the total
matter power spectrum. This indicates that on those scales 
the effect of baryons on the total matter power spectrum 
is dominated by the distribution of the baryons themselves.
Interestingly, in the transitional regime of $1\leq
k/(\mpch)\leq 5$, the differences between the \dmonly~power
spectrum and the DM component of \eagle~are typically as large as $\sim3\%$.
This is greater than the differences we observe in the
total matter $P(k)$. Consequently, even though  in this regime the effect
of baryons on the {\it total} matter power spectrum is small,
DM-only simulations will still fail to accurately
predict the power spectrum of the DM component.
Finally, at $k\geq10\mpch$ there is more power in \dmonly{},
than in \eagle{} DM, this reflects the fact that \dmonly{} simulations
cannot model depletion of gas from lower mass haloes caused by stellar feedback
and reionisation, which in turn makes virial masses of those haloes smaller
in hydro runs \citep[see \eg][]{Sawala2013,Schaller2015a}.
\begin{figure}
  \includegraphics[angle=0,width=83mm]{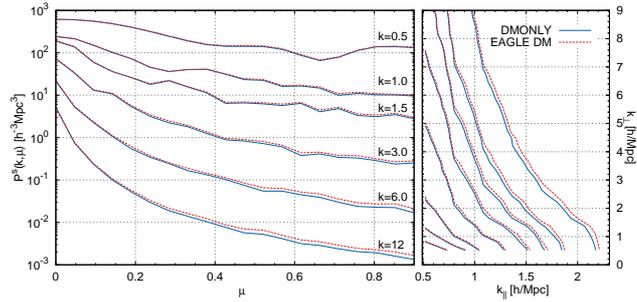}
   \caption{
    {\it Left panel:} the DM redshift-space power spectrum $P^s(k,\mu)$ computed at 6 different
    $|\vec{k}|$ intervals. {\it Right panel:} the full two-dimensional DM power spectra,
    different lines mark isoamplitude contours. 
    {\it Both panels:} the solid (dashed) lines correspond to \dmonly~(\eagle~DM) results.
     }
 \label{fig:rds_pk_compare}
\end{figure}

The effects that we have observed here for the DM and baryon
density fields are not surprising, considering all the
non-linear and highly energetic processes modelled by the
\eagle~simulation. The question that we
now want to answer is: to what extent and on what scales does the
non-linear physics
of the baryonic back-reaction induce changes on the velocity field?
We can do this by analysing
the red line in the bottom panel of
Fig.~\ref{fig:pk_compare}. This line depicts the absolute difference
between the amplitude of the DM velocity divergence power spectrum --
$P_{\theta\theta}(k)$, and that of the corresponding \dmonly{} simulation. 
The absolute difference is smaller 
than $3\%$ in the range $1\leq k/(\mpch)\leq 10$. At larger
scales the difference quickly drops below $1\%$ and at $k\sim
0.2\mpch$ it already becomes negligibly small ($<10^{-3}$).
Qualitatively and quantitatively similar behaviour is observed for the
density-velocity cross-power spectrum, where differences at $k<10\mpch$ are usually
smaller than those in $P_{\theta\theta}$. In the case of
the monopole of the redshift space power spectrum, $P_0(k)$,
the difference between \eagle~ DM and the \dmonly~result attains its
maximal value of $\sim 4\%$ at $k=4\mpch$; however, the baryonic back-reaction
drops below $1\%$ already for wavenumbers smaller than $0.5\mpch$. 
For the quadrupole, $P_2(k)$, at small scales ($k>3\mpch$) we observe
the effect of a similar size, while at large scales baryonic effects are even smaller.

Fig.~\ref{fig:rds_pk_compare} compares the full
two-dimensional (right) and fixed $|\vec{k}|$ intervals redshift \eagle{} DM and \dmonly{} power spectra.
For clarity we plot only isoamplitude contours of the full 2D spectra.
The \eagle~box size is probably too small
to allow for a proper modelling of large-scale modes ($k<0.1\mpch$) and the Kaiser
effect \citep{Kaiser1987} due to finite volume effects \citep[see][]{Colombi1994}.
However, 
the box is sufficiently large to appraise the impact
of galaxy formation on smaller scales, where the ``Fingers of God''
effect distorts the matter power spectrum
amplitude. The isoamplitude contours are systematically
shifted to higher $k_\perp$ values for \eagle~, hence
indicating that the back-reaction of baryons on the DM leads to
a slightly weaker suppression of small-scale power due to viralised motions inside
clusters and groups of galaxies. 
This effect 
can be better seen on the left panel,
where it is noticeable only for close to l.o.s.
directions (\ie{} $\mu>0.5$) and small scales $|\vec{k}|>1\mpch$.
We find that for $|\vec{k}|=1\mpch$ and $\mu>0.5$ the difference $|P^s_{DMO}/P^s-1|$ can typically be as large as $6\%$,
while at $|\vec{k}|=0.4\mpch$ it is contained below $1\%$ for the whole $\mu$ range.

We will discuss the
implications of our findings concerning the back-reaction of baryons onto
the DM density and velocity power spectra in the discussion section.

So far, with the exception of the total matter power spectrum, we have
focused on statistics derived from the velocities and positions of DM
particles in our simulations. These are not accessible with
astronomical observations but are used in theoretical modelling.
However \eagle~ also provides
catalogues of galaxies and the haloes they inhabit. This allows us to
compare the peculiar velocities\footnote{For the purpose of our analysis we define 
the galaxy/halo peculiar velocity as the velocity of its most bound DM particle.
 The centre-of-mass velocity definition gives consistent results.} 
of haloes
in the \dmonly~and \eagle~runs. By measuring these differences we
can assess the extent to which \dmonly{} simulations will suffer from halo
and galaxy velocity bias induced by ignoring baryons and their back-reaction
onto the DM and galaxy velocity field. Since baryonic physics affects the virial
masses of haloes \citep[\eg][]{Sawala2013}, comparing haloes at
fixed masses will suffer from the additional trend induced by
the changes in halo mass. To reduce this additional scatter, we first match haloes
between both runs, following the method of \citet{Velliscig2014}. 
For each halo in the \eagle~run we find its unique
counterpart in the \dmonly~run by identifying the structure that
contains the majority of its $50$ most bound particles. The same is
done for the \dmonly~halos and only pairs that can be matched
bijectively between the two simulations are kept in the catalogue
\citep[see also][]{Schaller2015a}. Having matched halo pairs between the
two simulations, we compute the difference between their respective peculiar
velocities and average this quantity in bins of both \eagle~
halo virial\footnote{For the virial mass we use $M_{200}$, \ie~ 
the mass contained in a sphere of radius $r_{200}$ 
centred on a halo, such that the average overdensity inside 
the sphere is $200$ times the critical closure density, $\rho_c$.}
and galaxy stellar mass. We find that
the averaged peculiar velocity difference is $\Delta |\vec{v_{p}}|\leq1$ km/s
for haloes with $M_{200}>3\times 10^{11}M_\odot$, hosting galaxies with
$M_\star>1\times 10^9M_\odot$. For haloes with galaxies more massive
than $M_\star\geq 3.5\times 10^{10}M_\odot$ the offset between the
\dmonly~and \eagle~halo peculiar velocities is consistent
with zero. 
The corresponding $1\sigma$ scatters are 20 km/s and 7 km/s respectively.
For all haloes the average difference is much
smaller than $\Delta |\vec{v_{p}}|$ of matched DM
particles, which is $\sim -4$ km/s, with $\sigma = 86$ km/s (\ie~ DM particles
in the full hydro run have smaller velocities). The average velocity 
differences are small, but the corresponding dispersions are larger.
We have checked that the bulk contribution to the quoted dispersions are coming
from large haloes and reflect the fact that differences in time integration between \dmonly{}
and \eagle{} run can capture a given particle at a different orbital position for the same
corresponding snapshot.
\citet{vanDaalen2014} have demonstrated that the difference in
the two-point correlation function of matched haloes
in the DMONLY and OWLS AGN simulations is negligible on scales larger
than the virial radius of the haloes.
In addition \citet{Schaller2015} have shown that vast majority of
\eagle{} galaxies show an offset between their luminous and dark matter
component that is smaller than the force resolution of the simulation.
A negligible effect on halo and galaxy velocities, that we find in \eagle{},
is thus consistent with their findings, as any long-lasting difference
in halo velocity would produce a significant position displacement over a Hubble time.

\section{Discussion}
\label{sec:conclusions}

We have measured and analysed systematic
differences in the DM density, velocity and redshift space power spectra between the full
\eagle~run and its dark matter only version at redshift $z=0$. 
The \eagle~ model of galaxy formation reproduces many properties of 
the galaxy population 
which suggests that the galaxy formation implementation 
is plausible in the sense that it does not invoke
unreasonably strong or weak feedback from star formation and AGN. This is
important as the work of VD11 showed that these two
processes mainly modulate the scale and strength of the baryonic back-reaction
onto the dark matter. However, recalling that \eagle{} 
overestimates the gas fraction in massive objects,
we can treat the results shown here as an approximate lower bound
on the magnitude of the baryonic influence on the DM.

Our findings imply that 
accurate modelling of hydrodynamical and galaxy formation
physics is essential to predict the total matter $P(k)$ on
scales corresponding to wavenumbers $k\sim 4\mpch(\lambda\sim1.6\hmpc)$ to better than $1\%$
accuracy. On larger
scales baryonic effects in \eagle{} change the amplitude by less than 
$1\%$, while on scales of $k\sim(3-6)\mpch[\lambda\sim(1-2)\hmpc]$ the change is greater than $10\%$.
This is a large number in the context of theoretical modelling of the
total matter power spectrum from weak lensing tomography
in forthcoming surveys such as Euclid or LSST 
\citep[\eg][]{Hearin2012}.
We stress that \eagle{} is expected to underestimate baryonic
effects since the cluster gas fractions are significantly too low S15.
This may explain the quantitative difference with VD11, who found a $1\%$
effect for $k>0.3\mpch(\lambda<21\hmpc)$ for the OWLS model AGN \citep{Schaye2010}
which does reproduce the observed gas fractions \citep{McCarthy2010}.

The amplitude of the power spectrum of the \dmonly~model deviates
by $\sim 3\%$ from the scaled DM component of the full \eagle~run
on scales of $1\simlt k/(\mpch)\simlt 5[1\simlt\lambda/(\hmpc)\simlt6]$.
This indicates that 
colisionless simulations fail to model 
the distribution of the DM component precisely. 
This was to some
extent already present in the results of \citet{Schaller2015a}, who
found that the DM density profiles of haloes that contain \eagle~galaxies
deviate from their \dmonly~counterparts. Our
results indicate that the DM distribution beyond the 
virial radii of haloes can also be significantly affected by the baryonic
back-reaction.

The impact of baryons on the DM peculiar velocity field is less
pronounced than on the density field, but it extents to
somewhat larger scales. Nevertheless, the effect seen in our simulations
is less than $1\%$ on scales $k\simlt 0.5\mpch$.  
This shows that baryonic
effects connected to the galaxy formation physics are not crucial 
to build accurate models of redshift spaces distortions, provided
that these models are restricted to sufficiently large scales. Since 
theoretical models of the shape and amplitude of the DM $P_{\theta\theta}(k)$
and $P_{\delta\theta}$ are the main ingredients of redshift space distortions
models \citep[\eg][]{Kaiser1987,Scoccimarro2004,Taruya2010,dlTorre2012},
it was important to appraise the magnitude and scales at which the baryonic
physics affects the expansion scalar power spectrum.

The impact of baryons on the peculiar motions of haloes and galaxies
is even smaller. This implies that baryonic effects are negligible
in the modelling of the large-scale velocity field of galaxies and haloes.
This is important because a number of  
velocity-based observables have been proposed to constrain cosmological parameters and models 
\citep[see \eg][]{Nusser1994,Strauss1995,Nusser2012,Tully2013,Hellwing2014,Koda2014}.

To conclude, our results suggest that DM-only simulations may be
sufficiently accurate to model the cosmic
peculiar velocity field of haloes, galaxies and dark matter.
However, baryonic effects are important 
and need to be taken into account in order
to attain the required accuracy of the total-matter and DM power spectra 
demanded by future surveys like Euclid or LSST.

\section*{Acknowledgements}
The authors thank the anonymous referee who helped improved
the quality of the paper. Peder Norberg, Shaun Cole, Maciej Bilicki, Adi Nusser, Enzo Branchini
and Agnieszka Pollo are also acknowledged for valuable discussions and comments.
WAH acknowledges support from the European Research Council grant through
646702 (CosTesGrav) and the Polish National Science Center under contract \#UMO-2012/07/D/ST9/02785.
This work was supported by the Science and Technology Facilities
Council (grant numbers ST/F001166/1 and ST/K00090/1); European Research Council (grant
numbers GA 267291 ``Cosmiway'' and GA 278594 ``GasAroundGalaxies'')
and by the Interuniversity Attraction Poles Programme initiated by the
Belgian Science Policy Office (AP P7/08 CHARM). RAC is a Royal Society
University Research Fellow. This work used the DiRAC Data Centric
system at Durham University, operated by the Institute for
Computational Cosmology on behalf of the STFC DiRAC HPC Facility
(www.dirac.ac.uk). This equipment was funded by BIS National
E-infrastructure capital grant ST/K00042X/1, STFC capital grant
ST/H008519/1, and STFC DiRAC Operations grant ST/K003267/1 and Durham
University. DiRAC is part of the National E-Infrastructure.

\bibliographystyle{mnras}
\bibliography{eagle_flows}
\bsp
\label{lastpage}

\end{document}